\documentclass[conference]{IEEEtran}

\usepackage{graphicx} 
\graphicspath{ {./Images/} }
\usepackage{float} 

\usepackage{amsmath}
\usepackage{amssymb}
\usepackage{mathtools}
\usepackage{booktabs}

\usepackage{caption}
\usepackage{refstyle}
\usepackage{subcaption}
\usepackage[style=base]{caption}

\usepackage{cite}

\usepackage{array}
\usepackage{wrapfig}
\usepackage{multirow}
\usepackage{tabu}

\usepackage{comment}
\usepackage[super]{nth}

\usepackage{algorithm2e}

\usepackage{tikz}
\usetikzlibrary{shapes.geometric, arrows}
\usetikzlibrary{matrix}
\usepackage{xcolor,colortbl}
\usetikzlibrary{calc}

\renewcommand{\vec}[1]{\mathbf{#1}}

\newcommand{\qrsDetectionText}{QRS Detection \\ \& \\ Segmentation}

\tikzstyle{arrow} = [thick,->,>=stealth]
\tikzstyle{process} = [rectangle, minimum width=2cm, minimum height=1cm, text centered, draw=black]

\begin{document}
\title{Continuous User Authentication using IoT Wearable Sensors}

\author{
\IEEEauthorblockN{Conor
Smyth\IEEEauthorrefmark{1},  Guoxin
Wang\IEEEauthorrefmark{1},  Rajesh
Panicker\IEEEauthorrefmark{2},  Avishek
Nag\IEEEauthorrefmark{3},  Barry
Cardiff\IEEEauthorrefmark{3},  Deepu John\IEEEauthorrefmark{3}}
   \IEEEauthorblockA{\IEEEauthorrefmark{1}\IEEEauthorrefmark{3}University College Dublin, Ireland,\IEEEauthorrefmark{2}National University Singapore, Singapore}
  \IEEEauthorblockA{Email: \IEEEauthorrefmark{1}{\{conor.smyth.2,guoxin.wang\}}@ucdconnect.ie,
  \IEEEauthorrefmark{2}rajesh@nus.edu.sg,
  \IEEEauthorrefmark{3}{\{avishek.nag, barry.cardiff, deepu.john\}}@ucd.ie}}
  
\maketitle

\begin{abstract}
Over the past several years, the electrocardiogram (ECG) has been investigated for its uniqueness and potential to discriminate between individuals. This paper discusses how this discriminatory information can help in continuous user authentication by a wearable chest strap which uses dry electrodes to obtain a single lead ECG signal. To the best of the authors' knowledge, this is the first such work which deals with continuous authentication using a genuine wearable device as most prior works have either used medical equipment employing gel electrodes to obtain an ECG signal or have obtained an ECG signal through electrode positions that would not be feasible using a wearable device. Prior works have also mainly dealt with using the ECG signal for identification rather than verification, or dealt with using the ECG signal for discrete authentication.
This paper presents a novel algorithm which uses QRS detection, weighted averaging, Discrete Cosine Transform (DCT), and a Support Vector Machine (SVM) classifier to determine whether the wearer of the device should be positively verified or not. Zero intrusion attempts were successful when tested on a database consisting of 33 subjects.
\end{abstract}

\begin{IEEEkeywords}
ECG, Continuous Authentication, IoT, Wearable, Dry Electrodes.
\end{IEEEkeywords}

\section{Introduction}
\label{ss:CA_intro}


User authentication is a key component in almost all cyber-physical systems.  Traditionally, this is performed discretely at the start of a session using either something the user is in the know of such as a password or pattern, or something the user is in possession of such as a key or card, or something performed by the user such as fingerprint or voice or a combination of these \cite{camara2018real}.  If the system positively verifies the user, it will remain open until either a manual log-out, a power loss or until a certain period of inactivity has elapsed. 

\smallskip
While such systems are currently dominant, several issues can be identified as being major security flaws, namely, if such a system is left logged-in and unattended, there is nothing to prevent an intruder from gaining access  \cite{niinuma2010soft}. Secondly, a genuine user can willingly, whether tricked or not, give an intruder unsupervised access to the system. Systems employing continuous user authentication, however, do not have such issues because, as soon as the user cannot be positively verified, the system would automatically log itself out.

\smallskip
While a large number of wearable biometric authentication systems exist \cite{blasco2016survey}, continuous authentication is a challenging task \cite{patel2016continuous}. A continuous authentication system would have to work in the background without disrupting the user's everyday activity \cite{acar2018waca} - for example, the user should not be asked to submit a fingerprint at regular intervals and also, the system should not log genuine users out mid-session. In this paper, it is proposed that the users would wear a chest strap, similar to a low-power heart rate monitor which one would wear while jogging \cite{NUSDavid, 23uW,9QRS}. The purpose of this chest strap would be two-fold. Firstly, the band contains an ECG monitor to record the ECG signal which will be used to constantly authenticate the wearer. Secondly, the individual sensors themselves could act as tokens, so through possession of a particular device, the system could have a further layer of security. These sensors could then communicate with other devices via a wireless communication protocol such as Bluetooth Low Energy (BLE), granting verified users access to other systems or facilities, etc. Such strategies are however outside the scope of this paper. Furthermore, to add yet another layer of security, a password at the start of a session on an end terminal could be used for the initial log-in, and then to remain logged-in, the wearer would have to be continuously authenticated by their particular sensor.

This paper deals with continuous verification as opposed to discrete identification. For discrete identification, the waiting time between the attempted log-in and a decision has to be kept short. This reduces the amount of time available to make a decision, and limits the amount of data that can be collected. Also, verification is a one-vs-one problem, which is simpler than the one-vs-many problem in identification. To this end, this paper proposes that subjects have their own unique binary classifier on their sensor which will be used to decide whether the wearer is the genuine owner of the sensor or an intruder. The wearer's ECG signal is first be segmented using the R-peaks and processed. This processing involves finding the weighted average of the past heartbeats found within the last $t_{avg}$ seconds, with the aim of reducing the intra-subject variability and reducing the effects of noise on the individual beats. This parameter $t_{avg}$ should be chosen to be long enough to give a significant improvement in classification performance but short enough so that if an intruder were to use the sensor immediately after a genuine user, most of the genuine beats would already be too \textit{old} to be in the weighted average. 

\smallskip
The weights are found by using an agglomerative hierarchical cluster tree to rank the heartbeats. This method was chosen as the Euclidean distance between noisy heartbeats and clean heartbeats will be larger than the distance between clean heartbeats. This means that noisy beats will have less of an impact on the average. Also, this method does not use prior knowledge about the owner's ECG signal which is important as prior attempts which used template beats from the owner, only served to increase the similarity between the heartbeats of intruders and the owner.

\smallskip
The DCT is then used for the purpose of data compression. This is necessary as the averaged heartbeats can be considered to have a high dimensionality and a classifier would perform as well in this situation. These DCT coefficients are used as the features for the classifier. This method was chosen as prior works \cite{second_order_stats} have had success using the DCT in the past for the purpose of dimensionality reduction in this field, and also because other works such as \cite{eigenPulse} suggest that using fiducial features alone is not an option as a significant percentage of the population have irregular heartbeats which makes enrollment of these individuals problematic.

\section{Database Acquisition}
\subsection{Need for a New Database}

As far as continuous authentication using an IoT wearable device is concerned, none of the known publicly available databases of ECG signals were deemed to be suitable. Here \textit{suitable} is understood to mean that:

\medskip
\begin{itemize}
    \item The signals are recorded using a \textit{genuine} IoT wearable device, i.e., not a holter monitor or any other medical device which has wires and/or gel electrodes.
    \item Each subject should have multiple recordings taken over several weeks.
    \item The length of each recording should be in the order of hours.
    
\end{itemize}
\medskip

To this end, a database was recorded using the MAXIM-ECG-MONITOR  \cite{maxim}. Thirty three subjects have taken part in the study, Eighteen of which have partaken in multiple sessions.

\subsection{Measurements}

Using the available APIs and source code from Movesense's example Android applications \cite{mobile_repository}, a new app was developed in order to log the ECG data from the subjects. The ECG signals were recorded at the highest possible sample rate of 512 Hz. Subjects were also asked to keep track of when they were sitting or doing other activities using the app but as this was expected to be error prone, measurements in the $x$, $y$, and $z$ directions from the sensor's accelerometer, gyroscope, and magnetometer were also recorded so that the activity levels of the subjects could also be monitored. Depending on the specific use-case, this could then be used to filter out sections of the ECG signals which correspond to periods of high or low activity, or be used to estimate the posture of the subjects. These measurements were recorded at 13 Hz. While wearing the sensor, subjects went about their daily activities such as working at a desk, etc. Note that although the MAX-ECG-MONITOR's user guide says that a gel or water should be placed on the electrodes to improve the quality of the signal, this suggestion was ignored as it was thought to take away from the \textit{wearability} of the sensor. Subjects have also been asked to fill out a survey form giving their age, sex, approximate height, and weight and if they have any known heart conditions. 


\section{Enrollment}

Many prior works claim that obtaining an enrollment signal longer than a few minutes would not be possible. This is a major issue because the ECG is a time-dependent signal which varies due to the subject's heart rate, posture, level of anxiety etc. This paper takes the view that short enrollment signals don't capture enough intra-subject variability to be used in continuous authentication.

\smallskip
Other works, \cite{perils, adaptive_strings, 24_hours}, suggest the use of template-updating so that the model always has an up-to-date record of the genuine user, however, this paper does not advocate for template-updating as this would require that the classifiers be periodically re-trained. This would likely entail sending the ECG signals to a server to be stored and analyzed which would pose risks to the subject's medical information. Also, there is no way to guarantee that a subject's model is not replaced over time with an intruder's biometric signature, assuming that their ECG signals were quite similar. For these reasons, long signals will be used for the enrollment stage. Obtaining long signals for enrollment is not seen to be an issue, as in continuous authentication, the users would be wearing the sensor for long periods of time anyway.

\smallskip
During enrollment, a model will be made for each subject. This model will contain a template heartbeat for the subject, minimum and maximum thresholds for the amplitude of the heartbeats, and a binary classifier to perform the verification. The templates are found by averaging many heartbeats together from the subjects enrollment signal. Likewise, the thresholds are chosen using the enrollment signal.

\begin{figure}
    \centering
    \includegraphics[width =0.9\linewidth]{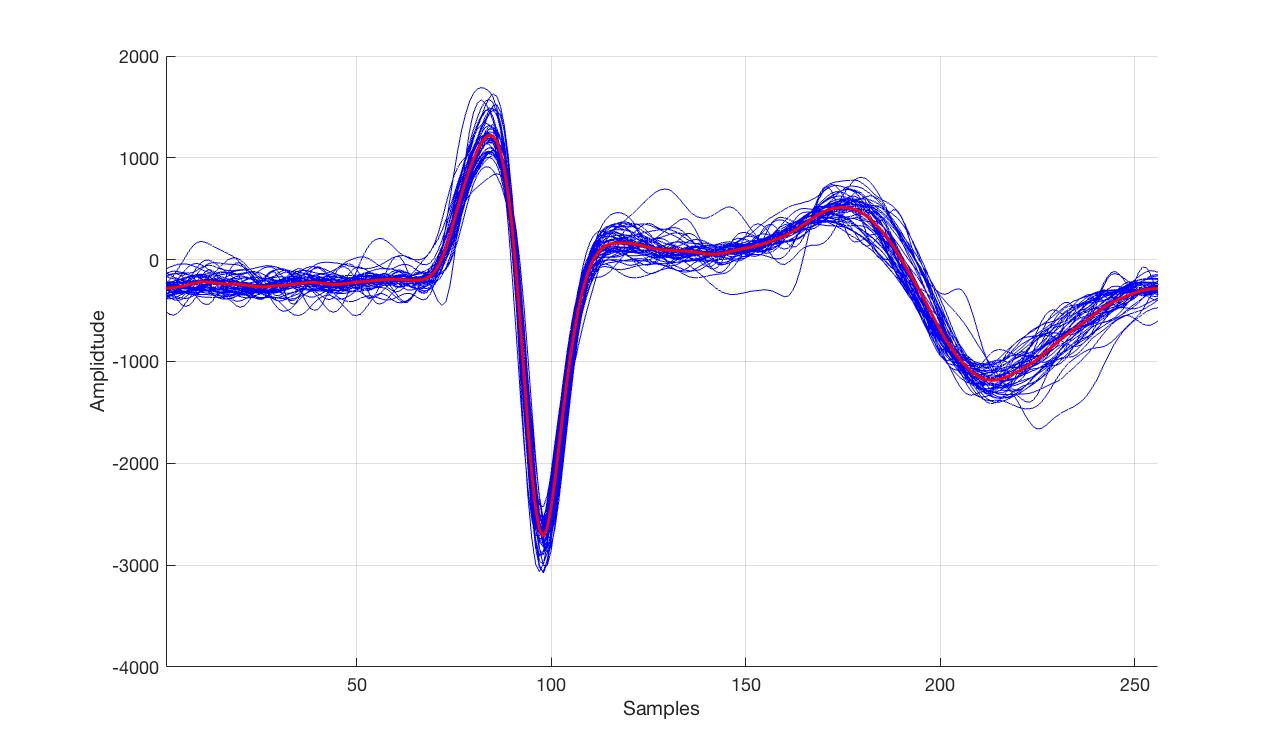}
    \caption{Heartbeats Found in a 30 Second window (Blue), and Their Average (Red)}
    \label{fig:avg_hearbeat}
\end{figure}

\section{Proposed Authentication Method}
\subsection{Processing}

Let $x[n]$ be the n$^{th}$ ECG sample obtained from the MAX-ECG-MONITOR, i.e., after the built-in baseline wander and powerline interference removal. These samples are fed into an online QRS Detector\cite{jointQRS}, which is a variant of the Pan Tomkins QRS Detector as described in \cite{Pan_tom}. This detector will return a vector, $\vec{X}_i \in \mathbb{N}^{N\times1}$ at time $t_i$. This vector will contain $N = 256$ samples, 78 of which are to the left of the detected QRS complex.

As this algorithm is prone to mistaking other sharp peaks as QRS complexes, the Pearson correlation coefficient (Eqn. \ref{Pearson}), $r_{xy}$, between this vector $\vec{X}_i$ and the owner's template heartbeat $\vec{Y}_i$ is found. Note that because the template beat is static, the template mean $\bar{y}$ and the standard deviation of the template $s_y$ can be pre-computed. If the correlation coefficient is below a certain threshold, $r_{min}$ the detected beat $\vec{X}_i$ is thought to be either belonging to the intruder or too noisy to be considered in the subsequent stages. This threshold $r_{min}$ has been set to 0.9 as it was empirically found to let few very noisy beats through. 

Heartbeats with a high enough correlation coefficient are added to a first in, first out (FIFO) buffer. This buffer contains all of the heartbeats found within the last $t_{avg}$ seconds. Each time a new heartbeat enters the buffer, the Euclidean distances between this new beat and the beats already in the buffer are calculated. Agglomerative hierarchical clustering is used to to cluster the beats in the buffer using the average linkage method. The order in which these beats are added to clusters is used to assign each beat a weight which are calculated using a Kaiser Window. This weighted averaging reduces the effect of noise, as can be seen in Fig. \ref{fig:avg_hearbeat}, which shows all of the time-warped beats found in a 30-second window in blue and their average in red.

\begin{figure}
    \centering
    \resizebox{0.9\linewidth}{!}{
    \begin{tikzpicture}[node distance=5cm, every text node part/.style={align=center}]

    \node (ECG)[xshift = -10cm]{\includegraphics[width=2.5cm]{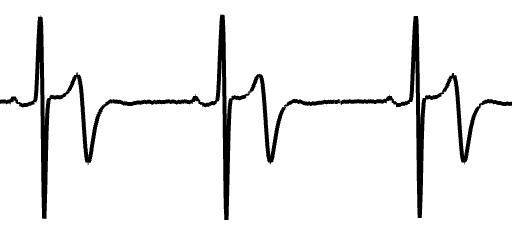}};
    \node (ECGText) [above of = ECG, yshift = -4cm]{ECG Signal};
    \node (QRS_Det) [process, right of= ECG,text width=3cm, xshift = -.5cm] {\qrsDetectionText};
    \node (Corr) [process,text width=2cm, below of = QRS_Det, yshift = +2cm] {Correlation Coefficient};
    \node (templates) [below of = ECG, , yshift = +2cm]{\includegraphics[width=2cm]{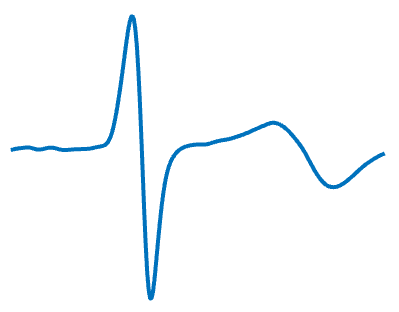}};
    \node (templateLabel) [above of = templates, yshift = -3.5cm] {Template \\ Heartbeat};
    \node (avg) [process, below of= Corr, yshift = +3cm] {Weighted \\ Heartbeat \\ Averaging};
    \node (DCT) [process, below of= avg, yshift = +3cm] {Discrete Cosine \\ Transform};
    \node (QDA) [process, below of= DCT, yshift = +3cm] {SVM \\ Classifier};
    \node (verf) [process, below of= QDA, yshift = +3cm] {Verification Decision};
    \node (status) [below of= verf, yshift = +3.5cm] {Log-in Status};
    \node (mdl) [left of = QDA, xshift = 2cm] {Model Parameters \\ (Computed Offline)};
    \node (ti_1) [left of = avg, xshift = 2cm, yshift = .25cm] {$t_i$};
    \node (ti_2) [left of = verf, xshift = 2cm] {$t_i$};
    \node (tavg) [below of = ti_1, yshift = +4.5cm] {$t_{avg}$};
    \node (measurement) [below of = QRS_Det, , yshift = +3.3cm, xshift = 2cm]{\includegraphics[width=2cm]{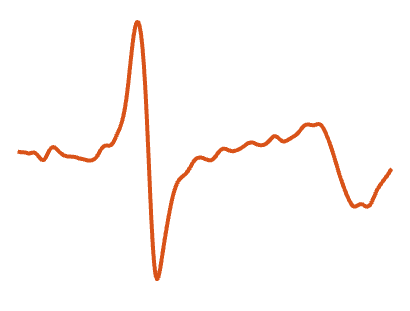}};
    \node (measurement) [below of = avg, , yshift = +4.4cm, xshift = 2.2cm]{\includegraphics[width=2cm]{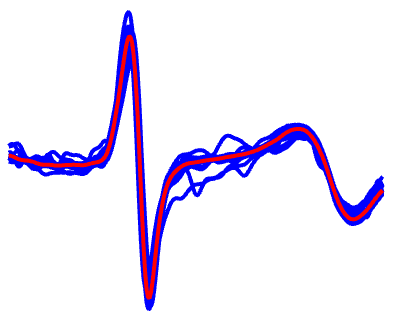}};
    \node (measurement) [below of = DCT, , yshift = +4cm, xshift = 2.5cm]{\includegraphics[width=2cm]{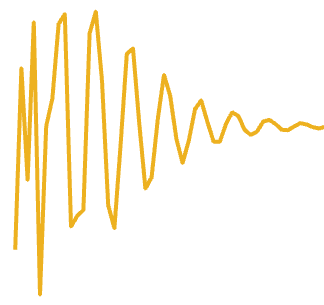}};
    
    \draw [arrow] (templates) -- node [anchor=south] {$\vec{Y} \in \mathbb{N}^{N\times1}$}(Corr);
    \draw [arrow] (ECG) -- node [anchor=south] {$x[n]$} (QRS_Det);
    \draw [arrow] (QRS_Det) -- node [anchor=east] {$\vec{X}_i \in \mathbb{N}^{N\times1}$} (Corr);
    \draw [arrow] (Corr) -- node [anchor=east] {$\vec{X}_i \in \mathbb{N}^{N\times1}$} (avg);
    \draw [arrow] (avg) -- node [anchor=east] {$\vec{A}_i \in \mathbb{R}^{N\times1}$}(DCT);
    \draw [arrow] (DCT) -- node [anchor=east] {$\vec{D}_i \in \mathbb{R}^{M\times1}$}(QDA);
    \draw [arrow] (QDA) -- node [anchor=east] {$z_i \in \{0, 1\}$}(verf);
    \draw [arrow] (verf) -- (status);
    \draw [arrow] (mdl) -- (QDA);
    \draw [arrow] (ti_2) -- (verf);
    
    \draw[arrow] (ti_1) -- node[above]{} ($(avg.180) + (0,1/4)$);
    \draw[arrow] (tavg) -- node[above]{} ($(avg.180) + (0,-1/4)$);

    \end{tikzpicture}
    }
    \caption{Block Diagram of Purposed Method During Verification Stage}
    \label{fig:block_diagram}
\end{figure}
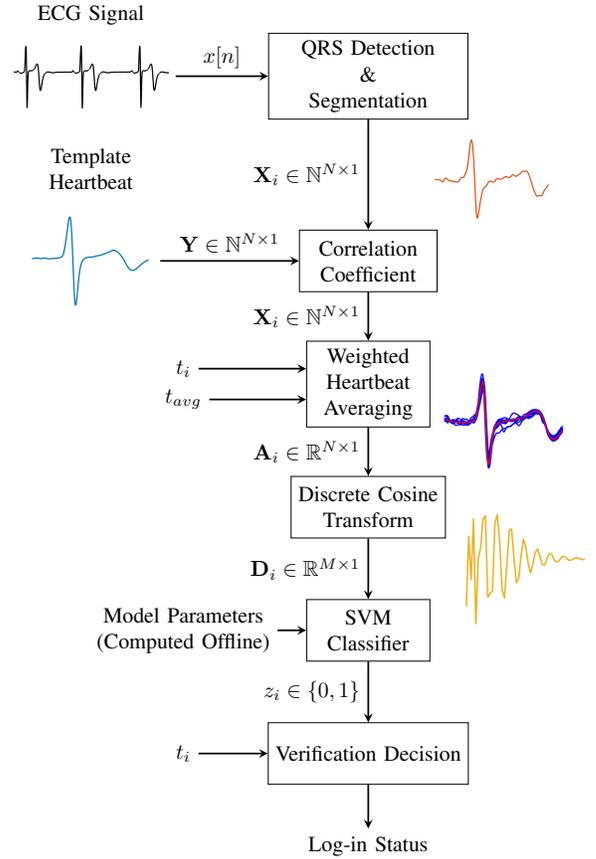

Finally, the weighted-averaged time-warped beat, $\vec{A}_i$  undergoes the Discrete Cosine Transform (Eqn. \ref{DCT_2_EQU}) for dimensionality reduction. This is desired as the classifier has better performance when less features are used. As the features, the first $M$ DCT coefficients are used, producing a vector $\vec{D}_i \in \mathbb{R}^{M\times1}$. Note that the terms after $x[n]$ can be pre-computed and stored in an $N \times M$ matrix.

\begin{equation}
\label{DCT_2_EQU}
y[k] = \sqrt{\frac{2}{N}}\sum\limits_{n=1}^{N} x[n]\frac{1}{\sqrt{1+\delta_{k1}}}\cos{\Bigg(\frac{\pi}{2N}(2n-1)(k-1)\Bigg)}
\end{equation}
\smallskip

\begin{equation}
\label{Pearson}
r_{xy} = \frac{\sum\limits_{n=1}^{N} x_i y_i - n\overline{x}\overline{y}}{(N-1)s_x s_y}
\end{equation}
\smallskip

\subsection{Classification}
A Support Vector Machine (SVM) classifier was used in this this experiment. SVM was chosen as in preliminary tests, it was shown to have both the lowest false positive rate and best Balanced Accuracy Rate (Eqn. \ref{BAR_EQU}) out of QDA, LDA, Na\"{i}ve, and Decision Trees. This is important as we want the proposed method to generalize well to unseen subjects. Note that this set of classifiers was chosen as an \textit{eager} learner classifier would be better suited to the problem over a lazy classifier such as KNN.

\subsection{Continuous Authentication}

Finally, in order to make the final decision on whether the wearer is in-fact the owner of the sensor, a simple threshold on the number of positive verifications $n$ obtained in the last $t_{v}$ seconds is enforced. 

\section{Results}

Subjects with only a single recorded session will either be used as intruders or will be enrolled into the system to make up a generic population, as described in \cite{blind_verification}. These subjects will not have their own classifier, as this would require splitting a same-day-session into two halves (one for training and one for testing), and in the authors' opinion, recordings taken from the same session should not be used for both enrollment and testing as this could give overly optimistic results.

\smallskip
Unlike most prior works, this paper is interested in accessing the performance of the classifier using subjects that the classifier has never seen before as intruders to the system. \textit{Unseen} intruders are used as, in the authors' opinion, it is not a fair assessment of the system performance if its only tested with subjects that it has seen before. To this end, a leave-one-out strategy was employed so that all of subjects with multiple sessions, have multiple classifiers trained for them, each one with a different subject left out to act as the intruder. 

\smallskip
As the number of observed heartbeats per subject in the test data varies drastically, the  Balanced Accuracy Rate, BAR (Eqn. \ref{BAR_EQU}), was used as the accuracy measure of the classifier as this metric is robust against skewed data. 

\smallskip
\begin{equation}
\label{BAR_EQU}
BAR = \Bigg(\frac{TP}{TP + FN} + \frac{TN}{TN + FP} \Bigg) / 2
\end{equation}

\begin{figure}
    \centering
    \includegraphics[width=0.9\linewidth]{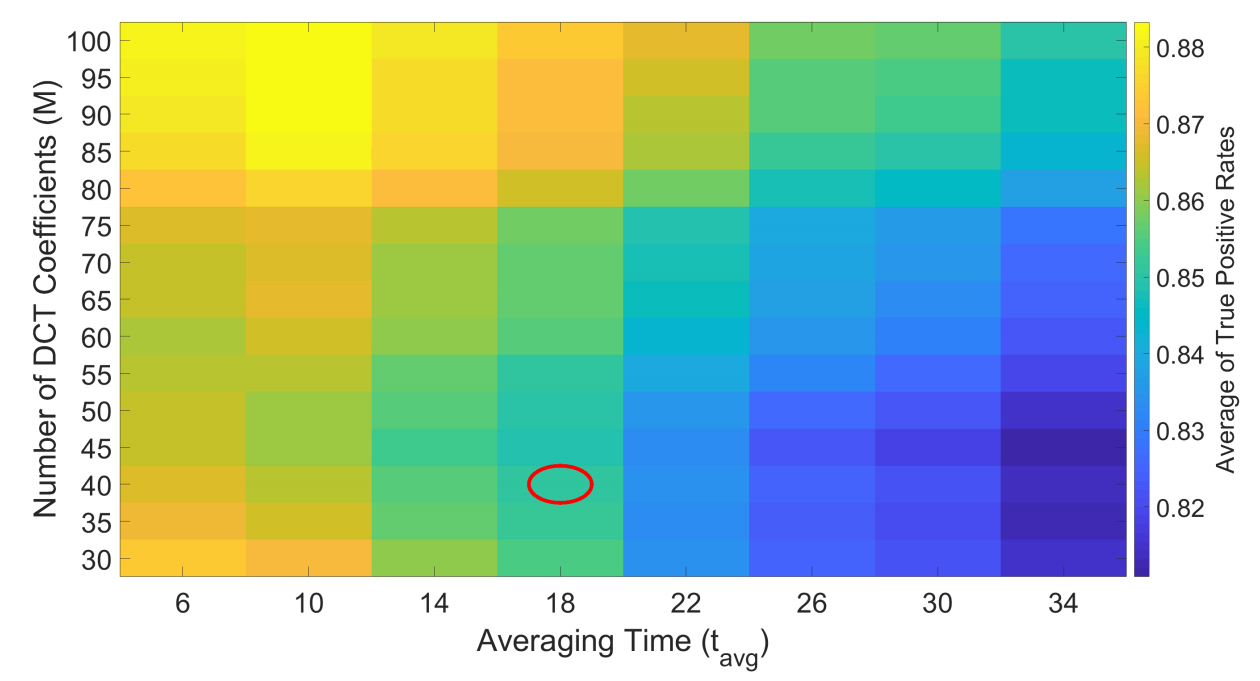}
    \caption{Parameter Determination Plots}
    \label{fig:tpr}
\end{figure}

\smallskip
The average BAR of all of the classifiers for each subject was then recorded as well as the BAR of the worst classifier. This evaluation method was used to test the performance of the system using different parameter values for $t_{avg}$ and $M$. Fig. \ref{fig:tpr} shows the best classification performance was obtained when $t_{avg} = 18$ seconds and $M = 40$.

\smallskip
These classifications are then used to calculate how often genuine users are logged out mid session and how often intruders and for how long intruders can gain access to the system. The time for which a genuine user is incorrectly logged-out is not relevant here - in a real life scenario, if a genuine user was logged-out, it would likely be due to the position of the strap which they could adjust. Likewise, special considerations should be made for cases where genuine subjects are constantly being logged-in and out as if they were aware that this was happening in a real-life situation, they could adjust the position of the sensor, change their posture, etc.

\begin{table}[]
\centering
\caption{Summary of Results from the Classification Phase}
\resizebox{0.9\linewidth}{!}{%
\begin{tabular}{c|c|c|c|c|c|c}
        & Sensor Owner       &         &         &         &         &        \\
Subject & Test Signal        & Average & Average & Average & Worst   & Worst  \\
ID      & Length {[}hh:mm{]} & BAR     & TRP     & FPR     & TRP     & FPR    \\ \hline
1704663 & 02:33:27           & 98.01\% & 96.01\% & 0.00\%  & 95.36\% & 0.00\% \\
1669218 & 01:29:24           & 99.48\% & 98.97\% & 0.00\%  & 98.80\% & 0.00\% \\
1757807 & 03:44:16           & 91.32\% & 82.63\% & 0.00\%  & 80.52\% & 0.00\% \\
1870493 & 01:03:47           & 98.36\% & 96.72\% & 0.00\%  & 96.21\% & 0.00\% \\
1744753 & 05:09:03           & 83.08\% & 66.16\% & 0.00\%  & 63.93\% & 0.00\% \\
0966761 & 05:30:02           & 87.75\% & 75.51\% & 0.01\%  & 72.97\% & 0.23\% \\
0974866 & 12:04:22           & 81.05\% & 62.11\% & 0.00\%  & 59.56\% & 0.00\% \\
1797219 & 14:53:22           & 83.73\% & 67.45\% & 0.00\%  & 65.05\% & 0.00\% \\
1652875 & 00:13:15           & 98.82\% & 97.64\% & 0.00\%  & 97.45\% & 0.00\% \\
1744114 & 03:58:22           & 64.43\% & 28.85\% & 0.00\%  & 22.89\% & 0.00\% \\
0976733 & 08:40:38           & 87.76\% & 75.53\% & 0.00\%  & 73.93\% & 0.00\% \\
1717488 & 08:40:08           & 87.25\% & 74.51\% & 0.00\%  & 69.41\% & 0.00\% \\
0897327 & 07:29:30           & 95.39\% & 90.79\% & 0.00\%  & 90.19\% & 0.00\% \\
1703978 & 02:28:11           & 78.19\% & 56.38\% & 0.00\%  & 51.14\% & 0.00\% \\
1855093 & 01:04:25           & 96.61\% & 93.21\% & 0.00\%  & 92.85\% & 0.00\% \\
1773663 & 01:15:05           & 98.93\% & 97.86\% & 0.00\%  & 97.60\% & 0.00\%
\end{tabular}%
}
\end{table}

\section{Conclusion}

This paper concludes that using an off-the-shelf IoT wearable device recording ECG is a viable option for continuous user authentication. As the data was mostly recorded while the subjects were working at desks, this form of continuous authentication is feasible in most office environments. This paper also acknowledges that there are concerns about the sensitive medical information that can be obtained from the ECG signal. Subjects have also expressed concern about being constantly monitored in a work environment. To alleviate these concerns, it is recommended that all records of the ECG signals be destroyed after the enrollment stage and that all processing of subsequent ECG signals be performed on the embedded sensor, so that the ECG data is not transmitted and stored anywhere.
\smallskip

It is planned to have the database expanded further by adding more subjects, as well as to add recordings of subjects doing a wide variety of tasks in order to test the limitations of using the ECG for continuous authentication. It is also planned that the length of time $t_{avg}$ over which heartbeats are averaged could be variable, up to a maximum of say 60 seconds. This length could be varied based on the activity of the subject, the signal quality of the heartbeats, etc. Also, it is believed that a weighted average of the heartbeats would give a better performance, so that noisy heartbeats have less of an influence on the average heartbeat. The length of time $t_{v}$ or the number of verifications $n$ required in this time could also be made variable as the number of verifications in a time period is very much dependent on the subject's heart rate.

\bibliographystyle{IEEEtran}
\bibliography{IEEEabrv,references}
\end{document}